# Persephone: A Pluto-System Orbiter & Kuiper Belt Explorer

Carly Howett[1], Stuart Robbins[1], Bryan J. Holler[2], Amanda Hendrix[3], Karl Fielhauer[4], Mark Perry[4], Fazle Siddique[4], Clint Apland[4], James Leary[4], S. Alan Stern[1], Heather Elliott[5,6], Francis Nimmo[7], Simon B. Porter[1], Silvia Protopapa[1], Kelsi N. Singer[1], Orenthal J. Tucker[8], Anne J. Verbiscer[9], Bruce Andrews[4], Stewart Bushman[4], Adam Crifasi[4], Doug Crowley[4], Clint Edwards[4], Carolyn M. Ernst[4], Blair Fonville[4], David Frankford[4], Dan Gallagher[4], Mark Holdridge[4], Jack Hunt[4], J. J. Kavelaars[10], Chris Krupiarz[4], Jimmy Kuhn[4], William McKinnon[11], Hari Nair[4], David Napolillo[4], Jon Pineau[12], Jani Radebaugh[13], Rachel Sholder[4], John Spencer[1], Adam Thodey[14], Samantha Walters[4], Bruce Williams[4], Robert J. Wilson[15], Leslie A. Young[1]

## Abstract

Persephone is a NASA concept mission study that addresses key questions raised by New Horizons' encounters with Kuiper Belt objects (KBOs), with arguably the most important being "Does Pluto have a subsurface ocean?". More broadly, Persephone would answer four significant science questions: (1) What are the internal structures of Pluto and Charon? (2) How have the surfaces and atmospheres in the Pluto system evolved? (3) How has the KBO population evolved? (4) What are the particles and magnetic field environments of the Kuiper Belt? To answer these questions,

[1]Southwest Research Institute, Boulder, CO 80302, USA; howett@boulder.swri.edu

[2]Space Telescope Science Institute, Baltimore, MD 21218, USA

[3]Planetary Science Institute, Tucson, AZ 85719, USA

[4]Johns Hopkins Applied Physics Laboratory, Laurel, MD 20723, USA

[5] Southwest Research Institute, San Antonio, TX 78228

[6] Physics and Astronomy Department, University of Texas at San Antonio, San Antonio, TX 78249, USA

[7]Department of Earth and Planetary Sciences, University of California Santa Cruz, Santa Cruz, CA 95064, USA

[8]NASA/Goddard Space Flight Center, Greenbelt, MD 20771, USA

[9]University of Virginia, Charlottesville, VA 22904, USA

[10]NRC Herzberg Institute of Astrophysics, Victoria, BC V9E 2E7 BC, Canada

[11]Department of Earth and Planetary Sciences and McDonnell Center for Space Sciences, Washington University, St. Louis, MO 63130, USA

[12]Stellar Solutions, Palo Alto, CA 94306, USA

[13]Department of Geological Sciences, Brigham Young University, Provo, UT 84602, USA

[14]Department of Systems Engineering, Colorado State University, Fort Collins, CO 80523, USA

[15]Laboratory for Atmospheric and Space Physics, University of Colorado, Boulder, CO 80303, USA

Persephone has a comprehensive payload, and would both orbit within the Pluto system and encounter other KBOs. The nominal mission is 30.7 years long, with launch in 2031 on a Space Launch System (SLS) Block 2 rocket with a Centaur kick stage, followed by a 27.6 year cruise powered by existing radioisotope electric propulsion (REP) and a Jupiter gravity assist to reach Pluto in 2058. En route to Pluto, Persephone would have one 50- to 100-km-class KBO encounter before starting a 3.1 Earth-year orbital campaign of the Pluto system. The mission also includes the potential for an 8-year extended mission, which would enable the exploration of another KBO in the 100- to 150-km-size class. The mission payload includes 11 instruments: Panchromatic and Color High-Resolution Imager; Low-Light Camera; Ultra-Violet Spectrometer; Near-Infrared (IR) Spectrometer; Thermal IR Camera; Radio Frequency Spectrometer; Mass Spectrometer; Altimeter; Sounding Radar; Magnetometer; and Plasma Spectrometer. The nominal cost of this mission is $3.0B, making it a large strategic science mission.

## 1. Introduction: The Case for Returning to the Kuiper Belt

NASA's New Horizons spacecraft blazed the trail of Kuiper Belt (KB) exploration with its encounter of the Pluto system in 2015 and close flyby of Arrokoth (2014 $MU_{69}$), a cold classical Kuiper Belt object (CCKBO), in 2019. Resultant spacecraft data led to several important discoveries: KBOs are diverse (see Figure 1), Pluto has a currently active surface, Charon has had an active geologic history, and the CCKBO Arrokoth is a contact binary. The data returned raised new questions that can be best answered by a return to the Pluto system with an orbiter, yet understanding the diversity of the KB and other dwarf planets also beckons. The Persephone mission would achieve both of these desires: to put an orbiter into the Pluto system and to explore the KB.

Images from New Horizons showed Pluto is unexpectedly active with vigorous surface geology, including a convecting ice sheet known as Sputnik Planitia (SP; Stern et al., 2015; McKinnon et al., 2016; Moore et al., 2016) filling an ancient basin. One possible explanation for the equatorial location of SP requires a subsurface ocean, sparking the debate about whether one could exist on Pluto (Nimmo et al., 2016), a body that spends most of its time >40 AU from the Sun. Models of Pluto's interior evolution show the maintenance of a subsurface ocean on Pluto is feasible (see references within Nimmo and McKinnon, 2021). Determining whether Pluto does indeed have a subsurface ocean is one of the key drivers for this mission because a subsurface ocean has important astrobiological implications for our solar system and, by extension, other systems as well (e.g., Hendrix et al., 2019).

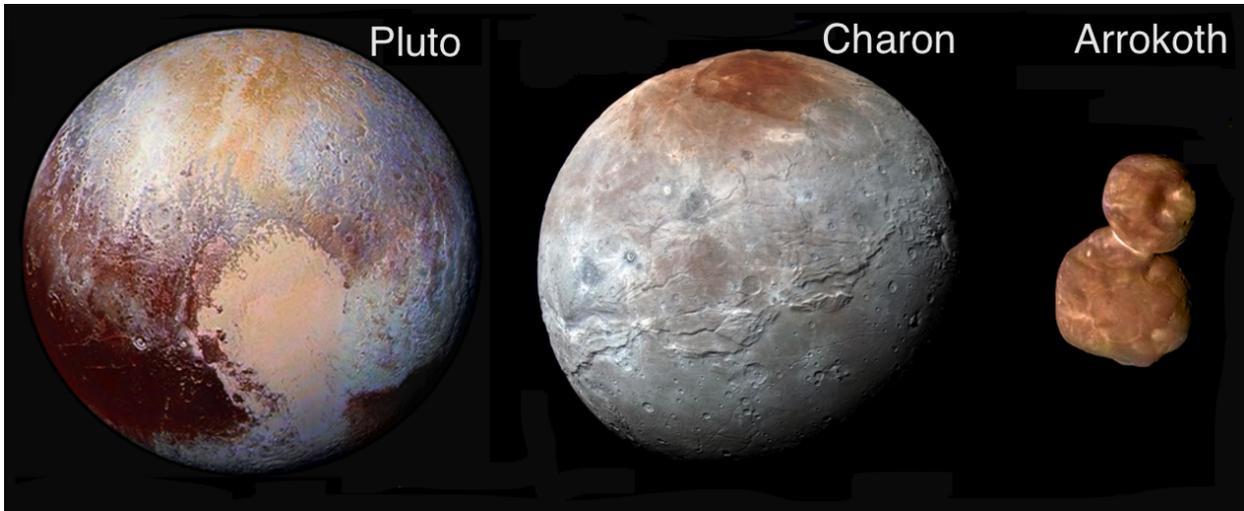

**Figure 1: The diversity of** explored **KBOs, from left to right: Pluto, its largest moon Charon, and the cold classical KBO Arrokoth. Note that these objects are not shown to scale: The diameters of Pluto, Charon, and Arrokoth are 2377 km, 1212 km, and 36 km (longest axis), respectively (Stern et al., 2015; Nimmo et al., 2017; Spencer et al., 2020). (Image credits: NASA/JHUAPL/SwRI.)**

Our detailed knowledge of other KBOs is lacking because of the difficulty in obtaining high signal-to-noise data for these faint, small objects from Earth-based facilities. We do know that the KBO population has diverse surface colors, albedos, and compositions, implying that KBOs are intrinsically different, and/or that they experienced different resurfacing processes (e.g., Barucci et al., 2008; Benecchi et al., 2019). Different KB regions appear to have different binary fractions, with the CCKBOs having the highest percentage of noncontact binaries (~30% compared with ~15% for the remaining populations; Noll et al., 2008, 2020) and plutinos having the highest (40%) fraction of contact binaries (Thirouin et al., 2018). CCKBOs having such a high fraction of binaries implies that their population is primordial, whereas the lower fraction of binaries of non-CCKBOs is consistent with a greater influence of dynamical evolution (gravitational scattering) and collisional processes (Parker & Kavelaars, 2012). Furthermore, the KB was assumed to be fully collisionally evolved like the asteroid belt for objects less than ~100 km across, but crater counts from New Horizons have called that assumption into question (Singer et al., 2019). Persephone's vantage point inside the KB provides a unique opportunity to resolve the surfaces of several KBOs and observe them at solar phase angles unattainable from Earth, which allows their photometric properties to be determined. The proposed mission would improve our knowledge of KBOs, KB binaries, and the evolution of the KB as a whole (Bernstein et al., 2004; Benecchi et al., 2018).

Understanding the nature of Pluto has profound implications for the evolution of other bodies in our solar system (e.g., Neptune's captured-KBO moon, Triton; Agnor & Hamilton, 2006) and would provide critical information on heat and volatile transport mechanisms in KBOs. Furthermore, understanding the diversity of the KB reveals its complex evolution and the context of the KB within the solar system's small body population.

We describe the science case for returning to the KB, and specifically discuss why the Pluto system is so compelling. A Pluto-system orbiter and KB explorer is a multi-decadal mission requiring many Next-Generation Radioisotope Thermoelectric Generators (NGRTGs). This type of mission (i.e., an orbiter rather than a single- or even multi-spacecraft flyby) was selected because the science return would be groundbreaking rather than incremental. It is important to point out that other mission architectures may result in decreased mission duration and cost—for example, another New Horizons-like spacecraft flyby of the non-encountered hemisphere. While any exploration of the KB would increase our knowledge of this enigmatic region, the science return from such a mission would be greatly diminished. For example, without an orbiter, it would be very difficult to answer conclusively whether Pluto has a subsurface ocean or fully understand the workings of Pluto's active geologic-climatologic engine. An orbiter would also allow the planetary plasma interactions with the solar wind and the interplanetary magnetic field (IMF), for example to determine the stability of the interaction boundaries and how they respond to changes both in the solar wind and IMF.

Instead of further exploring different mission architectures, which are explored by Robbins et al. (2020), we strongly support the development of new technologies that would decrease the risk of this mission. The most significant technological development would be the development of a nuclear electric propulsion system (NEP; cf. Casani et al., 2020), which would substantially decrease the cruise time to Pluto. Casani et al. (2020) show that a 10-kilowatt-electric (kWe) NEP spacecraft can deliver 67% more payload with 2.4 years shorter flight time compared to the current radioisotope electric propulsion (REP) system. They also showed that a kWe system would enable greater than four times the data rate at Pluto compared with the REP option.

We note that due to Pluto's obliquity (~122°), latitudes between 50°S and the south pole on both Pluto and Charon would be in shadow during the mission. However, many of the surface-sounding instruments (WAC, thermal IR camera, altimeter, and sounding radar) will still be able to sense the dark surface.

The name Persephone was chosen for the mission after the queen of the underworld in Greek mythology. Given that Pluto is named after the Roman god of the underworld, and that we wanted a female name to reflect our diverse team with many women in leadership roles, this name seemed apt.

## 2. Science Objectives

The high-level science objectives for the Persephone mission can be broadly separated into four categories: (1) constrain the internal structures of Pluto and Charon, (2) reveal surface and atmosphere evolution, (3) provide context for KBO formation and evolution, and (4) explore the particles and magnetic field environments of the Kuiper Belt. These objectives would be achieved by using the proposed instrument payload, which includes multiple imagers and spectrometers, a mass spectrometer, an altimeter, and a radar sounder (§3), while following the proposed mission design, including pre- and post-Pluto KBO encounters (§4). Figure 2 gives an overview of the scientific questions that motivate these objectives. Figure 3 presents maps with labeled features on both Pluto and Charon to orient the reader.

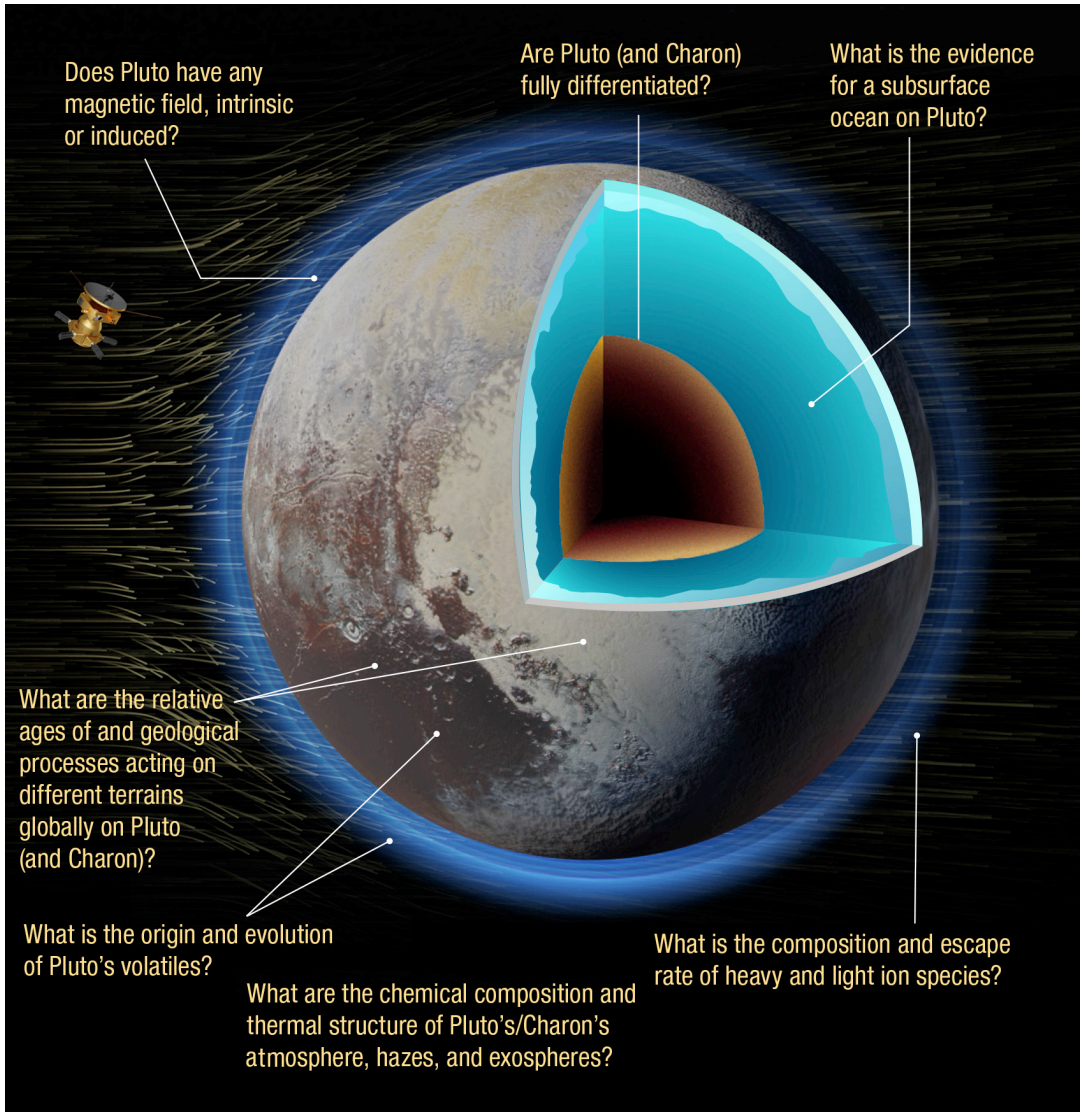

**(a) Science Questions of the Pluto-system**

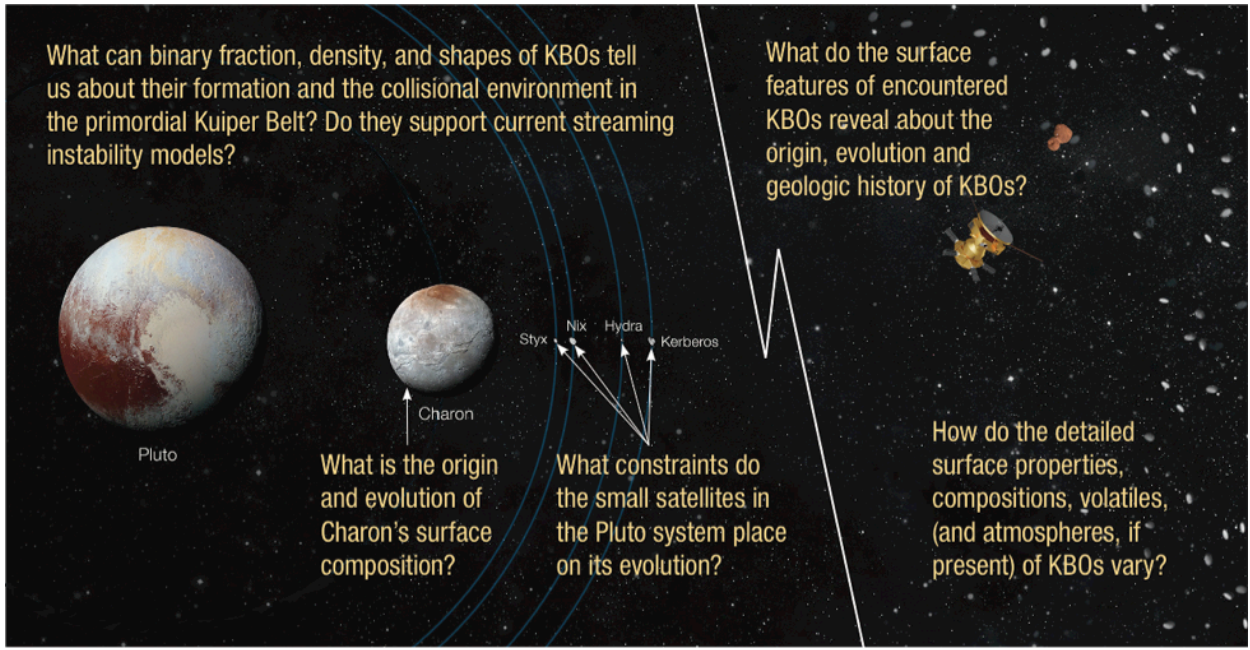

**(b) Science Questions that address wider KBO Science**

**Figure 2: Driving science questions addressed by the Persephone mission.**

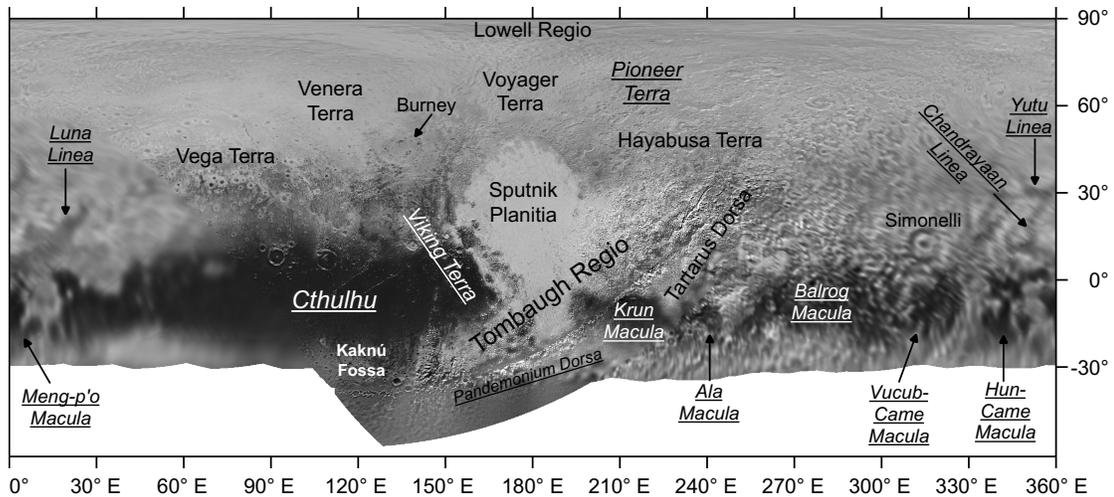

**(a) Pluto mosaic with feature names.**

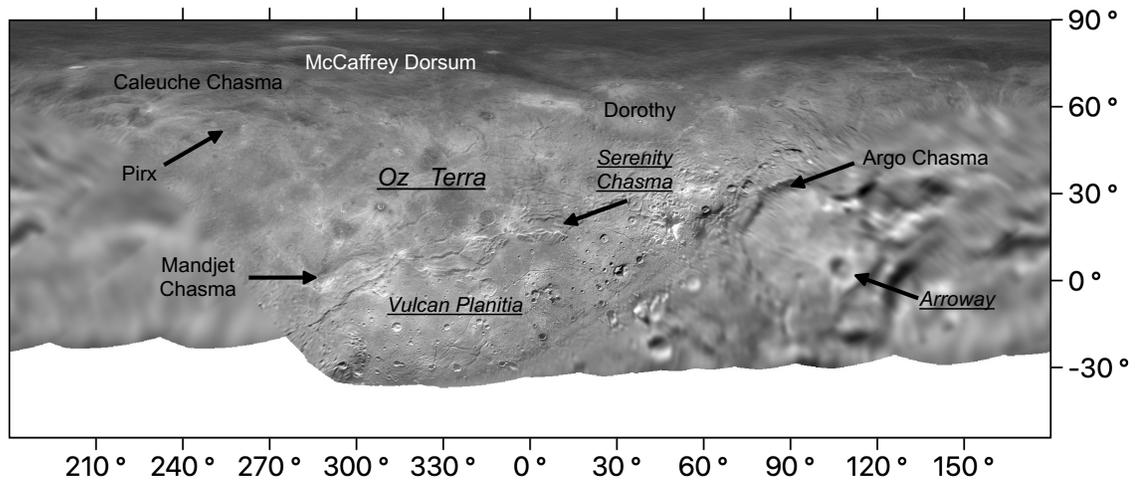

**(b) Charon mosaic with feature names.**

**Figure 3 - Maps of Pluto and Charon with features labeled, IAU-accepted names are given in regular font. Underlined names in italics are informal names. From Beyer et al. 2021.**

*2.1. Constrain the Internal Structures of Pluto and Charon*

Pluto's bulk density indicates that its internal composition is roughly two-thirds rock and one-third ice by mass (McKinnon et al., 2017), with an uncertain carbonaceous component, but how the rock and ice are distributed is unclear (Figure 4). The reason the distribution is important is that the extent to which rock and ice/water have separated can tell us about how much heat was released as Pluto accreted, and thus how accretion and evolution proceeded. The absence of compressional features indicates Pluto is not a homogeneous rock-ice mixture (McKinnon et al., 2017) but rather a closer to a partially or fully differentiated body: it might also have a Titan-like hydrated rock core. Meanwhile, Charon's smooth plains and vast tectonics are evidence of early heating, potential melting, and global expansion as the interior cooled and expanded (Beyer et al., 2019; Robbins et al., 2019), indicating that partial or full differentiation also may be possible for Pluto's largest companion. Pluto and Charon's differentiation states can be inferred from their moment of inertia (MoI). If the shape has relaxed to that of a fluid body, then either the present-day rotational or tidal bulges or the equivalent gravity coefficient ($J_2$ and $C_{22}$) could be used to deduce the MoI. No sign of "fossil bulges" was detected by New Horizons (Nimmo et al., 2017), but the limits derived are not very stringent (<0.6%, or 7 km).

The internal structure of Pluto is particularly intriguing because it could currently host a subsurface ocean. Such an ocean could help to explain how SP formed: if the basin was created by an impact, Pluto could have reoriented to its current position from tidal and rotational torques. These torques require the basin to be a positive gravity anomaly (despite being negative topography), which is best explained (from shell thinning and ocean uplift; Nimmo et al., 2016) if Pluto has a subsurface ocean. The heat within such an ocean could be insulated and maintained by a thin layer of clathrate hydrates (Kamata et al., 2019). However, other possible explanations for SP's formation do not require a subsurface ocean. For example, Hamilton et al. (2016) argue that SP was formed by the natural accumulation of ice at latitudes 30°S/N because they are Pluto's coldest regions. The only way to determine whether Pluto does indeed harbor an ocean is to return to the Pluto system with an orbiter so that its gravity and activity signatures can be determined.

Persephone would determine the internal structures of Pluto and Charon using a technique similar to that applied to Saturn's moon Enceladus, whereby a combination of global gravity and topography measurements determine the MoI (which gives differentiation state) and potentially ice shell thickness (Iess et al., 2014). Two-way Doppler radio science measurements would determine the gravity field along with long-wavelength topography obtained through a combination of laser altimetry and stereo imaging. Constraining Pluto and Charon's interior structures using these observations requires knowledge of their global shape with an accuracy of <100 m. Furthermore, accurate regional topography across areas of interest, particularly SP, needs to be known to <100 m. If a subsurface ocean is present, then a substantial gravity anomaly is predicted from SP (Nimmo et al., 2016). Radio Science Doppler tracking during multiple close flybys over a range of latitudes and longitudes would determine the gravity coefficients $J_2$ and $C_{22}$ of both targets. The magnetometer would be used to search for evidence of an induced or intrinsic magnetic field around Pluto, to determine whether subsurface exploration by induction is feasible for isolated KB worlds.

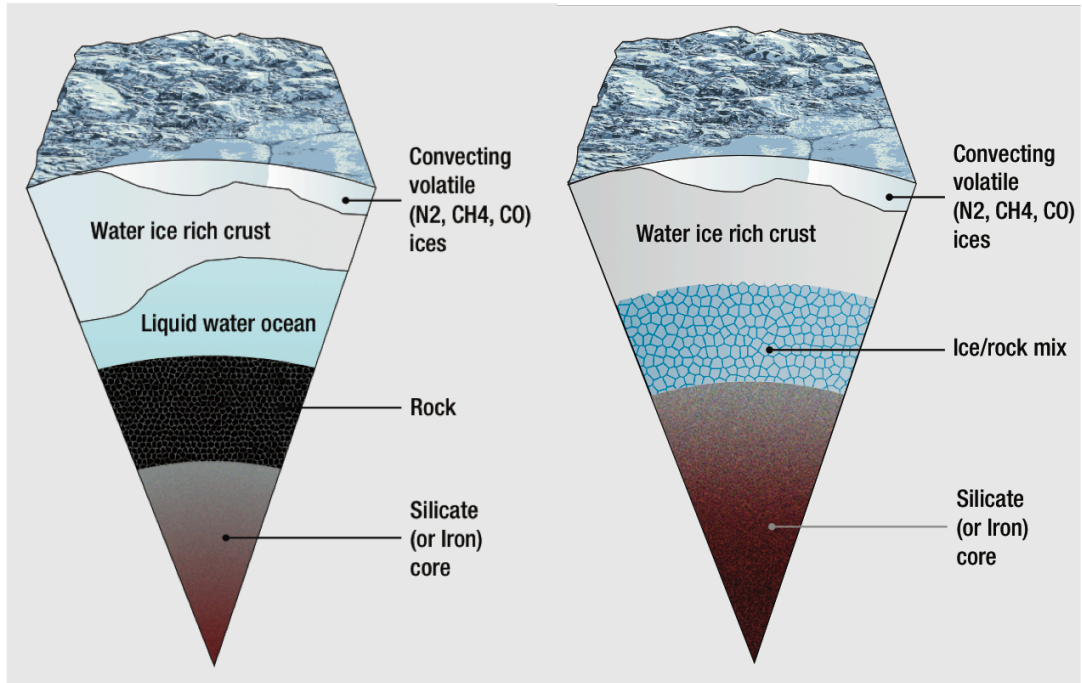

**Figure 4 – Possible scenarios of Pluto's interior.**

Pluto's rock-to-ice ratio is large enough to power low levels of internal heat for the duration of solar system history (Robuchon & Nimmo, 2011; McKinnon et al., 2017). However, it is surprising that Pluto displays features indicating the transfer of this internal heat to the surface (e.g. those shown in Figure 5) when larger bodies (such as Calisto) do not. Mapping all occurrences of water ice-based cryovolcanism across Pluto, with estimates for the ages from impact craters, would enable locating the source of these eruptions and measuring how much volume is erupted over time. Convection patterns in SP would also be examined, looking for any differences between the 2015 New Horizons observations and those from Persephone in ~2058, along with radar sounding through the depth of the SP ice sheet to estimate the heat required to drive the speed of convection and thereby obtain an estimate for global heat flow. Mapping Pluto's emission at long infrared (IR) wavelengths would enable estimation of its heat flow and searches for hot spots or other thermally anomalous regions associated with activity (e.g., Spencer et al., 2006). We note that the expected surface temperatures on Pluto are between 35 and 55 K (Earle et al., 2017).

By studying other large KBOs Persephone would be able to determine whether Pluto's activity is typical behavior for KBOs. The rock-to-ice ratio could be inferred for other KBOs prior to a flyby via characterization of a possible satellite orbits and the physical properties of the system using Earth-based telescopes.

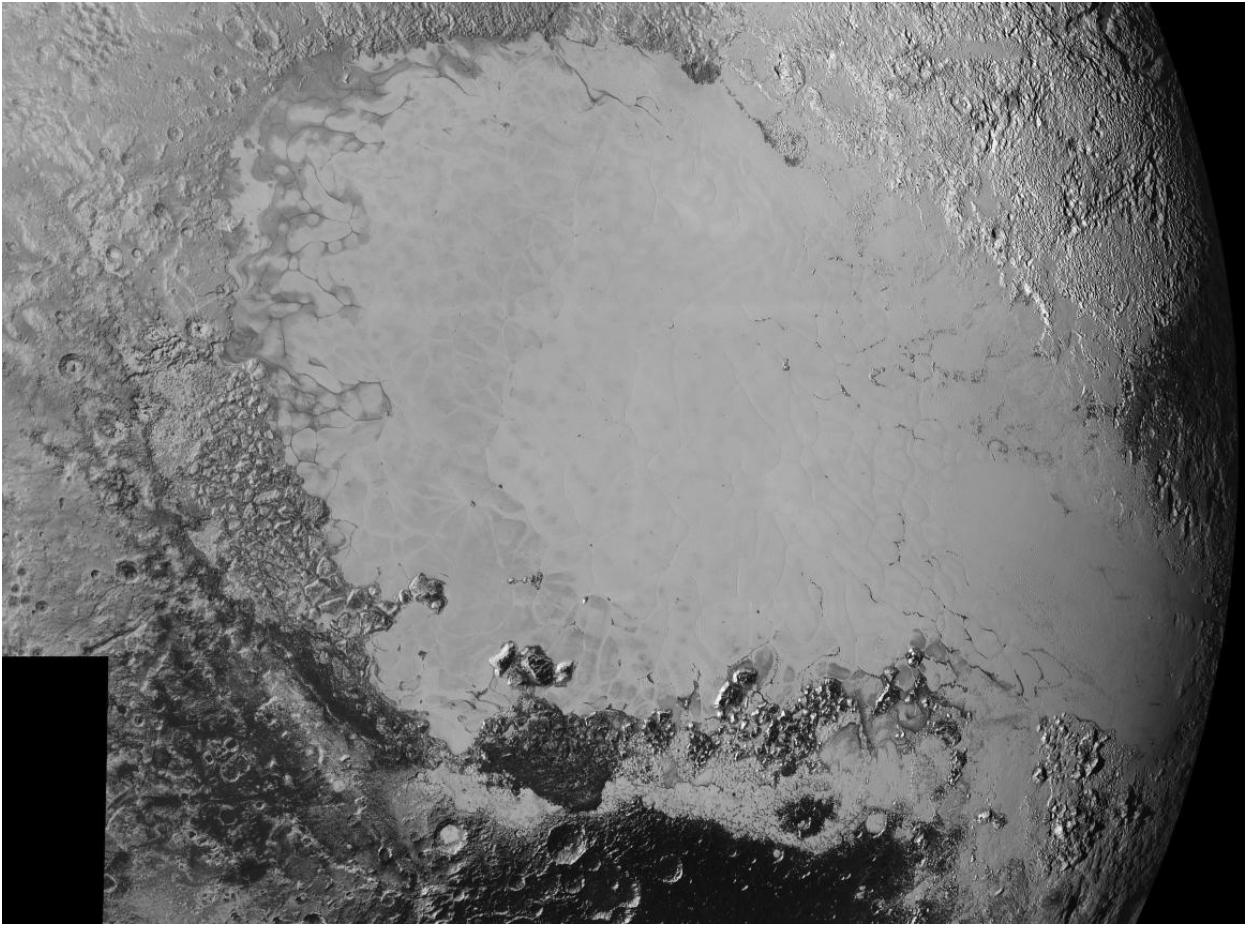

**Figure 5 (a) Details of Sputnik Plantia on Pluto as seen by New Horizons. Each of the irregularly-shaped segments are believed to be an overturning cell. This mosaic covers ~1,600 km (1,000 miles) and the smallest visible features are 0.8 km (0.5 miles) in size. (PIA19936, NASA/JHUAPL/SwRI)**

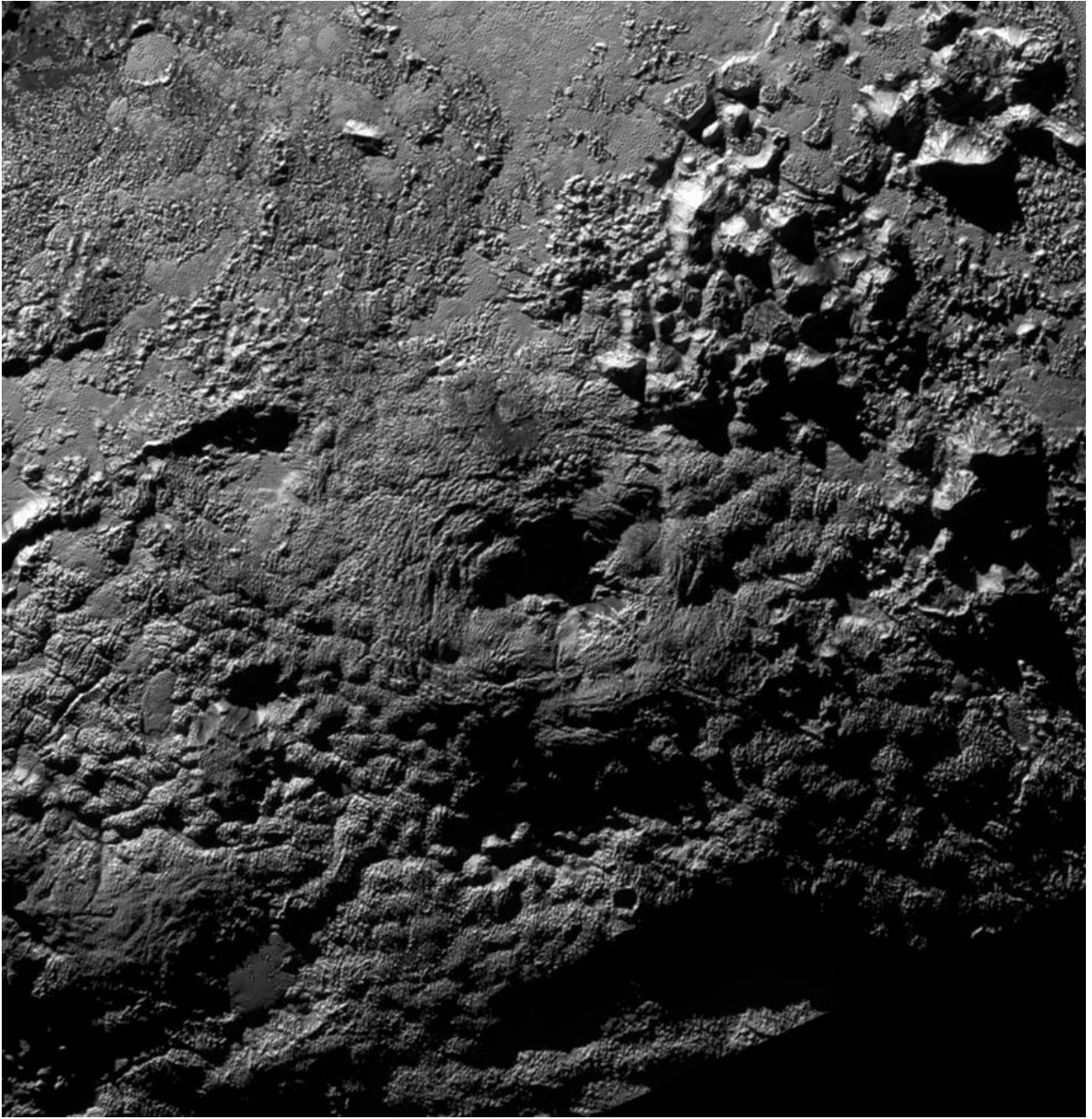

**Figure 5 (b) Wright Mons, a 160 km (100 mile) wide  possible cryovolcano on Pluto, as seen by New Horizons (PIA20155, NASA/JHUAPL/SwRI).**

**Figure 5 – Images of features consistent with both past and present activity on Pluto**.

*2.2. Reveal Surface and Atmosphere Evolution*

Both endogenic and exogenic processes have sculpted the surfaces of Pluto and Charon. The creation and degradation of geologic features reveal information about the interiors of both bodies, including their heat flow history and records surface-atmosphere interactions over time. Pluto and Charon both exhibit geologic features that are unique to those bodies.

It is possible that Pluto was even more internally active in the past than it is now. There are indications of surface expressions of internal activity, such as two possible cryovolcanoes, Wright and Piccard Montes (Fig. 5b) (Moore et al., 2016; Singer et al., 2018), and there is large-scale extensional fracturing across Pluto's equatorial belt that is consistent with freezing of a subsurface ocean. Furthermore, the red coloration and spectral signature of ammoniated ice on the extensional tectonic structures east of SP, known as Virgil Fossae, may have been due to the eruption of $NH_3$-$H_2O$ and colored "tholin-like" organics (Cruikshank et al., 2019). Dalle Ore et al. (2019) argue that since ammonia is geologically short-lived if exposed to solar UV radiation and charged particle bombardment, this implies it was either geologically recently deposited or uncovered.

Pluto's surface ices manifest in part as a large, continent-sized sheet of nitrogen ice, flowing nitrogen ice glaciers, mountains capped in methane ice, and extensive bladed terrain features that appear to be controlled by sublimation (Moore et al., 2018). Each landscape is dominated by exotic ice, and thus has different morphological appearances. The nitrogen glaciers appear to have relatively low viscosities and flow around blocky, hard water ice mountains, while the methane-bladed terrains, snows and related ices appear to be harder and more resistant, based on the fact that they can make dunes and maintain steeper topography. Better understanding these ices is important for this concept mission and requires measuring the slopes, and searching for changes over time (from New Horizons to ~2058) and determining exotic ice material properties at Pluto temperatures, such as rigidity, viscosity, brittle strength, and internal friction.

A big difference between Pluto and Charon is that Charon's optical surface is more porous (Verbiscer et al., 2019). Another difference is that because most geological activity occurred early (Singer et al., 2019), the lithospheric thickness recorded may be quite different from the present-day value (as at Mars). Charon's Vulcan Planitia is a remnant of early internal activity, and its smooth surface is thought to be due to an early subsurface ocean that erupted and resurfaced the area via ammonia-rich cryovolcanism (Beyer et al., 2019); altimetric and compositional mapping of Charon's surface composition would enable such hypotheses to be tested. Gravity and topography data would be used in admittance analyses, which provide both the elastic (lithospheric) thickness and the density of the near-surface material (e.g., McGovern et al., 2002). If portions of Charon's shell experienced foundering as a result of density contrasts, associated gravity anomalies should be present. Gravity data would be obtained by radio science measurements, and topography would be obtained by stereo imaging and radar altimetry. We could also search for and map smaller putative cryovolcanic constructs for regions of recent (or evidence of ongoing) activity across all of Charon's surface. Understanding both these targets (Pluto and Charon) would broaden our knowledge of how geologic processes operate on a broad range of icy worlds.

We would also model the surface ages of major regions of Pluto and Charon through cratering ages and crosscutting relationships. Furthermore, impact craters that formed at different times may show topographic and gravitational signatures of different degrees of relaxation, potentially probing the thermal history of the lithosphere, as at the Moon and the icy satellites (Bland et al., 2012; Kamata et al., 2015). These analyses can be achieved by high-spatial resolution color and panchromatic surface images.

Understanding the detailed surface properties of Pluto's minor satellites Styx, Nix, Kerberos, and Hydra may offer key constraints on the Charon-forming impact and subsequent accretion process (Figure 6, Canup, 2005, 2011; Stern et al., 2006). High-resolution images of the minor satellites would greatly enhance knowledge of the cratering record on their surfaces. Pluto's minor satellites have more craters than CCKBO Arrokoth despite it being comparable in size to Nix and Hydra (Robbins et al., 2017; Singer et al., 2019; Spencer et al., 2020; and see below). Because the poles of Pluto's minor satellites precess at unknown rates, it is hard to predict their orientation when the spacecraft arrives (Showalter & Hamilton, 2015); thus their local time coverage is impossible to predict. However, all instruments able to observe the winter-dark side of Pluto will able to observe the dark-side of the satellites too (e.g. WAC, altimeter, THEMIS, radar). By combining high-resolution images and altimetry, the shapes and volumes of the minor satellites can be determined (Weaver et al., 2016). Combining this result with dynamical estimates from radio science of the masses of the minor satellites would enable measurement of their bulk densities (Brozović et al., 2015; Porter et al., 2017) and hence likely composition and porosity; for instance, a density in excess of 1 g cm$^{-3}$ would require a significant rock fraction and/or a reduced porosity, which would make it unusual given knowledge of Saturn's smaller icy moons.

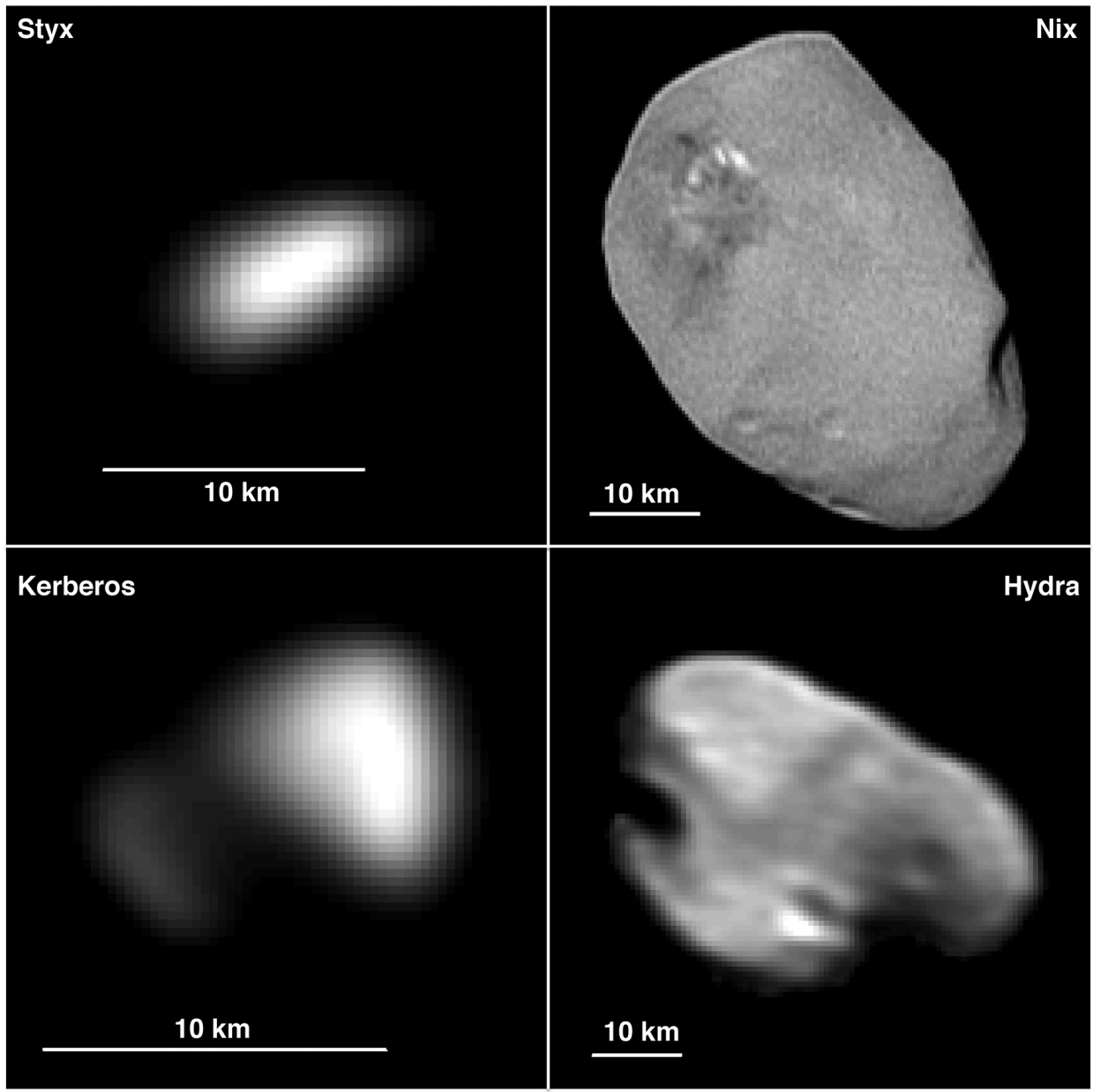

**Figure 6 – Highest spatial resolution images of Pluto's small moons as seen by New Horizons (from Weaver et al., 2016). Nix and Hydra are shown at their original pixel scales: 0.3 km/pixel and 1.97 km/pixel respectively. Styx and Kerberos have been deconvolved from their original pixel scales (3.13 and 1.97 km/pixel respectively) and resampled to higher pixel scales (0.39 and 0.25 km/pixel) for cosmetic reasons.**

New Horizons obtained the first resolved infrared spectra on three of Pluto's minor satellites: Nix, Hydra and Kerberos – showing the presence of both the 1.5 and 2.0 µm water ice band on all three satellites (Cook et al., 2018). Furthermore, bands of crystalline water ice and ammoniated species were detected on Nix and Hydra. No evidence was found for longitudinal differences in the disk-averaged spectra of Nix or Hydra. Nix's 2.21 µm crystalline water ice signature reaches a minimum in the spatial region associated with red coloring (Cook et al., 2018). Persephone would obtain high-signal-to-noise ratio (SNR), spatially resolved spectra for all four minor satellites, allowing direct measurements and comparisons of their surface compositions. Thus, (assuming they can be distinguished from the background) it would also enable an estimate of how much of their surfaces (and the surface of Charon) are contaminated with inter-satellite impact ejecta exchange (Stern, 2009; Porter & Grundy, 2015). Measuring the crater population on the smallest bodies within the Pluto system provides an important control on Pluto and Charon's surface age analysis because the small bodies should not be affected by the resurfacing on the larger bodies.

The study of geologic processes is closely linked to composition, which we know reasonably well on large scales. However, the composition of Pluto at sub-kilometer scales is unknown, because New Horizons mapped its composition to ~3 km/pixel on the encounter (and 70 km/pixel on the non-encounter) hemisphere. However, the color of Pluto (for example, across the bladed terrain of Tartarus Dorsa) varies over sub-kilometer distances, implying the composition might as well (Moore et al., 2018). High-spatial-resolution (<1 km/pixel) mapping would enable the composition of previously unexplored terrains to be determined. The composition of Pluto's darker surface materials, such as those that dominate Cthulhu, is still a mystery because New Horizons was unable to probe the diagnostic 2.5- to 5.0-µm spectral region. It is thought that the materials are organics, but exactly what type and how the material compares to those seen on other KBOs (e.g., Charon's North polar region) are unknown. Therefore, mapping Pluto's composition at high spatial resolution (<1 km/pixel) over a longer wavelength range than New Horizons (i.e., up to 5.0 µm) is key to understanding the composition and evolution of Pluto's surface. The short-wavelength cutoff of our proposed spectrometer is 0.8 µm, which is at shorter wavelengths than New Horizons'

LEISA (1.25 μm). The long-wavelength cutoff is also extended to 5 μm (from 2.5 μm) to differentiate between organic species (e.g., alkenes, alkanes) and tholins and to detect fundamental ice absorptions.

New Horizons' observations of Pluto's surface composition and atmosphere showed how the two are inextricably connected: sublimation and condensation of $N_2$ create a kilometer-deep daily piston of cold $N_2$ gas (referring to the thickness of the boundary layer), directly detected by New Horizons at dusk over SP via radio occultation, and likely are the cause of the layering imaged in Pluto's global photochemical haze layer (Cheng et al., 2017; Hinson et al., 2017) or orographic forcing (wind blowing over topography) (Gladstone et al., 2016). Both sublimation and orographic forcing can drive gravity waves, which maybe present in Pluto's atmosphere (Gladstone et al., 2016). Over Pluto's 248-year orbit, changes in solar insolation due to eccentricity and obliquity govern the surface pressure (currently 11 μbar) of its primarily $N_2$ atmosphere, which is in vapor pressure equilibrium with the surface ice (e.g., Meza et al., 2019). Upon arrival in 2058, we expect Pluto's surface atmospheric pressure to be lower than it was in 2015, but still global in extent, and likely higher than at the time of discovery of the atmosphere in 1988, based on post-encounter models and ongoing ground-based stellar occultations (Olkin et al., 2014; Meza et al., 2019). Thus, studying Pluto's atmosphere in particular by performing the first direct determination of its composition through mass spectrometry, remains a key objective.

In situ measurements of Pluto's atmosphere by a mass spectrometer would characterize the extent of volatile loss and transport with measurements of the global and temporal variations of $N_2$, $CH_4$, $C_2H_x$, and the myriad photochemical products that have not been measured, such as $H_2$. The minimum altitude for operating the mass spectrometer is expected to be ~500 km, below which saturation may affect the results based on New Horizons atmospheric densities. This altitude is consistent with the predicted density peaks of photochemical products and ions produced from the absorption of solar UV/Extreme UV (EUV) (Krasnopolsky, 2020). UV and radio occultations, UV spectra, mass spectrometry, and ion data would be obtained to understand the dynamics of the neutral and charged particle environment within Pluto's atmosphere, and to determine the composition and densities of Pluto's ionosphere (if one exists) (Hinson et al., 2018). Ultimately this information would provide an insight into the complex thermal balance of the atmosphere, key to understanding the extent of atmospheric loss during Pluto's lifetime.

Temporal variability in Pluto's surface composition and atmospheric composition and structure would be monitored in two ways: (1) by comparing the change between 2015 (when New Horizons encountered the system) and Persephone's arrival in 2058 and (2) through monitoring change throughout Persephone's ~3-Earth-year orbital tour of the Pluto system. In this way, changes over multi-decadal, yearly, monthly, and even hourly timescales could be probed.

Over much longer (~3-4 Myr) time periods often referred to as mega-seasons, cycles of Pluto's precession have influenced global volatile transport about its surface and volatile loss (Stern et al., 2017; Bertrand et al., 2018). Volatile loss from Pluto is governed by solar heating of the upper atmosphere due to the absorption of solar ultraviolet/extreme ultraviolet (UV/EUV) radiation primarily by $CH_4$. Pluto's upper atmosphere is much cooler (~70 K instead of ~90 K) than expected for a $N_2$-$CH_4$-dominated atmosphere, but the principal cooling agent affecting the thermal balance of energy in the upper atmosphere remains an enigma (e.g., Young et al., 2018). Nevertheless, the atmosphere is currently undergoing significant volatile loss via thermal escape due to Pluto's low gravity. In fact, light, minor species, such as $CH_4$, populate Pluto's extended corona, and a small fraction of the volatiles are eventually shared with Charon (e.g., Hoey et al., 2017; Grundy et al., 2016; Tucker et al., 2015).

The dark poles of Charon are thought to be cold-trapped volatiles ($CH_4$) from Pluto's extended atmosphere ($CH_4$) photolytically processed into more refractory material (Grundy et al., 2016). It is expected that periods of significant volatile loss and atmosphere transfer to Charon can occur when Pluto and Charon are near perihelion, governed by increased UV/EUV heating of the upper atmosphere. Evidence of volatile loss and the origin of Pluto's volatiles can be inferred from the isotopic abundances (e.g., $^{14}N$/$^{15}N$, $^{12}C$/$^{13}C$, D/H, and $^{40}Ar$/$^{36}Ar$) of the atmosphere (Mandt et al., 2017; Glein & Waite, 2018). Measuring the composition of Charon's poles in the 2.5- to 5.0-μm region, and the isotopic abundance of Pluto's atmosphere through mass spectrometry, would enable this hypothesis to be tested.

New Horizons placed upper bounds on the densities in Pluto's ionosphere; however, its composition remains unknown (Hinson et al., 2018). With a Pluto orbiter, we can probe the dynamics of the neutral and charged particle environment within Pluto's atmosphere. Ultimately, these observations would inform the complex thermal balance of the atmosphere, essential to understanding the extent of atmospheric loss during Pluto's lifetime. Furthermore, in the unlikely event that Pluto's atmosphere is contracted then the ion plasma observations would provide critical composition information. New Horizons found a long heavy ion tail, and these plasma measurements can quantify the escape rates for specific species. Additionally, an orbiting mission with plasma measurements quantifies the shape and extent of the solar wind - Pluto interaction, and quantifies how the interaction boundaries and tail vary in response to changes in the solar wind and interplanetary magnetic field.

## 2.3. Provide Context for KBO Formation and Evolution

Small KBOs are some of the most un-processed bodies in the solar system. Increasing our sample size beyond the one flyby of Arrokoth, which revealed a primitive world from the era of planetesimal formation (Lacerda & Jewitt, 2007; Stern et al., 2019), would greatly enhance our understanding of the diversity of KBOs and the processes that affect them. The shapes of other cold classical KBOs could reveal further clues to the formation mechanisms active in the primordial KB, and examination of their cratering records would reveal their subsequent interactions. Arrokoth has a somewhat uncertain crater population because of the illumination at flyby, although it appeared to have relatively few impacts relative to similar-aged asteroids, or even relative to Nix (Singer et al., 2020; Spencer et al. 2020). Different units across Pluto have different crater spatial densities because of resurfacing over time. Charon appears to have a mostly ancient surface (with early resurfacing in the equatorial smooth plains), but it also does not have a heavily cratered surface relative to more familiar inner solar system bodies, or even in comparison with Saturn's moons. Understanding craters in the KB has significant implications for the evolution of the KB as a whole. This is because measuring the crater population via imaging data, on a variety of terrains from all visited bodies, would give us numerous sample points. These data then inform the impactor population that created them. This analysis can be performed for different KB locations and object types, and potentially in time (because of different terrain ages on Pluto and Charon).

Hubble Space Telescope observations are able to separate KBO binaries separated by several thousands of km (e.g. Veillet et al., 2002). Similar remote observations, but with a different perspective will be possible from Persephone while in orbit around Pluto and during the pre- and post-Pluto cruise. Pluto will be located near the inner edge of the classical KB in the 2050s, thus only 2-10 AU from classical KBOs; this much closer proximity than from Earth would help resolve tighter binaries separated by only a few thousand km. Deep searches for binaries and faint satellites would be possible and important, and determining a satellite's orbit would ultimately lead to system masses and densities, informing models of binary formation in the primordial KB. Time-domain imaging would provide well-sampled rotational light curves, providing clues to KBO shapes and upper limits on density for those not in binary systems (as has been done for asteroids; e.g. Pravec et al., 2002)

Images obtained by Persephone would enable the production of maps of photometric properties over a large range of wavelengths and phase angles in order to understand Pluto's global surface variation. These could then be directly compared to both Earth-based and derived photometric properties of other KBOs by New Horizons (Porter et al., 2016; Verbiscer et al., 2019) and Persephone to compare their surface properties. Spectra of volatile-poor regions such as Cthulhu Macula on Pluto and Mordor Macula on Charon would provide useful baselines for understanding the red colors of many small KBOs that are challenging to observe spectroscopically from Earth.

*2.4. Explore the Particles and Magnetic Field Environments of the Kuiper Belt*

New Horizons provided the first in situ particles and fields data of the Pluto system. However, it did not carry a magnetometer and so was only able to indirectly characterize Pluto's magnetic field environment. Sending a magnetometer and plasma spectrometer would, for the first time, allow mapping the extent of large KBOs' magnetic environment (like Pluto's and Charon's), including the bow shock, "Plutopause" (interaction boundary between the solar wind and Pluto), and Pluto's ion tail. Pluto's bow shock is estimated to be located to be 4 between 5 $R_p$, based on both modeling and data from New Horizons' Solar Wind Around Pluto (SWAP) instrument (McComas et al., 2016; Feyerabend et al., 2017). Thus, this aspect of the mission would be explorative for Pluto and other visited KBOs, providing data about their interiors, surface modification processes that influence their evolution, and (where feasible on non-Pluto KBOs) atmospheric escape. If the 25-day solar wind variability is large enough, it might be possible to deduce a liquid ocean by modeling the induced magnetic field perturbations it would cause under different phases of this 25-day cycle. Including such instrumentation also makes this mission cross-disciplinary between planetary science and heliospheric science at NASA.

The required sensitivity of the plasma spectrometer is $m/\Delta m > 8000$ to enable $N_2$ to be distinguished from CO. Its energy range reproduces full coverage of New Horizons' Solar Wind Around Pluto (SWAP; McComas et al., 2008) instrument range and the low end of New Horizons' Pluto Energetic Particle Spectrometer Science Investigation (PEPSSI; McNutt et al., 2008) where most detections were near the bottom of its range (PEPSSI saw nothing at its high electron range; Bagenal et al., 2016). This sensitivity would resolve $H^+$, $He^+$, $He^{++}$, $O^+$, and $N^{+2}$ and thus properly characterize the solar wind's interaction with Pluto's atmosphere, possible ionosphere, and possible magnetic field.

## 3. Spacecraft and Instrument Payload

The Persephone design consists of a single spacecraft powered by five NGRTGs and carrying 11 unique science instruments. Persephone was designed with a long (>50-year) mission in mind and thus includes a variety of hardware redundancies. At the time of writing, the Voyager 1 and 2 spacecraft had been operational for 43 years; no mission yet executed has ever reached the 50-year milestone, let alone been planned from the outset for such a long operational lifetime.

*3.1. Spacecraft Design*

The Persephone spacecraft would be a three-axis controlled platform making use of three reaction wheels at any one time. Two additional reaction wheels would be left in reserve. Reaction wheels are typically designed for 15-year lifetimes on geosynchronous satellites, but some have lasted 20 years or more. The reaction wheels would be rotated in and out of use over time to extend their lifetimes due to their crucial role in maintaining spacecraft pointing during science observations.

For power generation, Persephone would make use of 5 NGRTGs mounted in a ring around the propulsion section (Figure 8) and thermally isolated from the rest of the spacecraft (though waste heat would be utilized to keep the xenon tank from freezing). This large number of NGRTGs is required to power the Radioisotope Electric Propulsion (REP) system and ensure usable power levels for instrument operation at the end of the extended mission, which is estimated to be <50% of the initial power level of 2000 W. The REP system would be used for the low-thrust cruise to Pluto, with a chemical propulsion system for higher-thrust maneuvers, such desaturating the reaction wheels and instrument pointing when necessary. Xenon, stored in a 4000-kg capacity tank, would be ionized and electromagnetically accelerated through 3 thrusters by the REP system to provide thrust to the spacecraft. The chemical propulsion system would make use of hydrazine and 16 thrusters for high-thrust maneuvers. Since these thrusters are powered by pressure in the hydrazine fuel tank, use of the chemical propulsion system and depletion of the tank would cause the specific impulse to decrease over time.

The NGRTGs would also power the suite of communications instruments**,** which includes two low-gain antennas, one medium-gain antenna, and one high-gain antenna (HGA). The latter would be 3 meters in diameter, similar to that on Europa Clipper, and would be the antenna used for communicating with the 34-meter Deep Space Network (DSN) arrays on Earth. Both X- and Ka-bands would be utilized, with the Ka-band providing 28 kbps, equating to 800+ Mb/day of data downlink while in Pluto orbit and ~350 Mb/day during the extended mission.

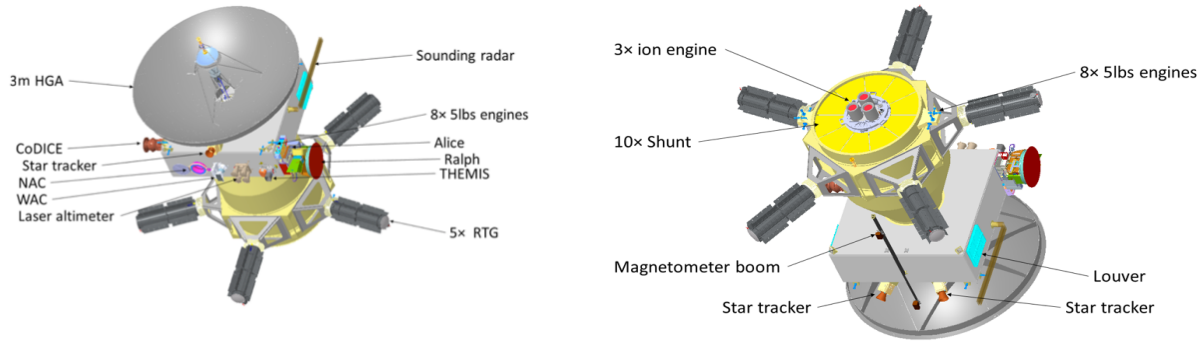

**Figure 8: (Left) Persephone external spacecraft overview featuring payload locations. (Right) Persephone external equipment layouts. Note the pentagonal distribution of the NGRTGs and the titanium brackets for thermal insulation.**

*3.2. Science Instrument Payload*

The entire science payload is based on existing instruments. None requires new technologies, but some have modifications that improve their performance for this mission. All of the instruments are mounted to the spacecraft or to a fixed boom. The imagers, imaging spectrometers, and altimeter are mounted on the same side of the spacecraft and would face nadir during most low-altitude passes. A perpendicular side of the spacecraft faces toward the direction of motion; the in situ instruments— magnetometer, plasma spectrometer, and mass spectrometer—are mounted on this deck. None of the instruments articulates, and the spacecraft rotation angle provides instrument pointing. This design prevents the need to anticipate and plan for a potential failure of a scan platform.

Most of the imaging and spectroscopic instruments can achieve the spatial resolution required to answer the science questions from altitudes of 5000 to 10,000 km. Because the spacecraft would dwell in this altitude range for thousands of hours over the mission lifetime, the instruments can perform their global mapping from this range. There is a range of illumination and viewing geometry available, which enables high-quality imaging and photometric studies. The observation plan requires only slow slews and low acceleration, well within the capabilities of the reaction wheels. The in situ instruments operate primarily within 2000 km of the surface but would also be used for several surveys of the Pluto/Charon system environment. The altimeter and radar sounder gather data at altitudes <1200 km. The science instruments are presented in more detail below; names in parentheses correspond to the labels used in Figure 8.

***Panchromatic and color high-resolution imager (NAC):*** The panchromatic and color imager provides high-spatial resolution global maps for panchromatic, eight colors, and stereo images for a total of 10 global maps. Measurements would be performed with different illumination and viewing geometry for phase-function calibration. The instrument would operate in a push-broom mode for Pluto and Charon, capturing both panchromatic and color bands simultaneously with color filters on eight strips that are 4000 pixels wide. There is also a framing mode for optical navigation and for long-range imaging of KBOs; this instrument serves the same purpose as LORRI on New Horizons (Cheng et al., 2007). The spacecraft provides the pointing and the 50 µrad/s scan rate. The pixel size is 5 µrad (~1"/pixel) for all data, the same as LORRI. Spatial resolution is <50 m for altitudes <10,000 km, and the surface scans occur between 7000 and 11,000 km. Spatial resolution for color data is lower, 200 m.

***Low-light camera (WAC):*** Due to Pluto's obliquity (~122°), latitudes between 50°S and the south pole would be in shadow during the mission. The low-light camera uses reflected light from Pluto to image the Pluto-facing hemisphere of Charon, and Pluto's winter hemisphere (using scattered light from Pluto's atmosphere and the minimal Charon-light) The low-light camera is a version of the Europa Imaging System (EIS) WAC (Europa Clipper; e.g., Turtle et al., 2014).

***UV spectrometer (Alice):*** The UV spectrometer would measure Pluto's atmosphere by providing column density versus wavelength during solar and stellar occultations. There are two apertures. The one for stellar occultations is aligned with the nadir-facing instruments and can also be used for characterizing surfaces of the airless targets (Charon, the minor satellites, and other KBOs). A second aperture has a restricted throughput designed for solar occultations and can be used at the same time as RF spectrometer measurements of occultations with Earth transmissions. This instrument is modeled after the Alice spectrometer onboard New Horizons (Stern et al., 2008).

**Near-IR spectrometer (Ralph):** The near-IR spectrometer provides hyperspectral coverage from 0.8 to 5 µm, an extension of the wavelength coverage of both New Horizons and Lucy LEISA (Reuter et al., 2008; SwRI 2020). The spectral regions are selectable, with ~250 of the 2000 spectral elements stored and transmitted to Earth. This capability enables high spectral resolution for important features across a wide range of wavelengths. The instrument is capable of angular (spatial) resolutions higher than 60 µrad.

**Thermal IR camera (THEMIS):** The thermal IR camera provides temperature measurements of a target's surface. The camera is based on the Mars-observing THEMIS instrument (Mars Odyssey; Christensen et al., 2004). Pluto's surface temperature is between 37 and 45 K (Earle et al., 2017), which means it's blackbody temperature emission peaks at ~130 µm. This is much longer than the peak of Mars' blackbody emission curve, which is at ~42 µm, due to its warmer (~210 K) surface. Thus we require different wavelength filters from THEMIS to observe Pluto's thermal emission, and baseline those of Diviner (Lunar Reconnaissance Orbiter; Paige et al., 2010). Measurements at several local times provide information on the thermal inertia of the surface.

**Radio Science (3M HGA):** The radio science experiment has two distinct aims: One is radio science, specifically to measure the degree-2 spherical harmonics of Pluto's mass distribution for studies of internal structure and planet formation. The second is to work as a radio frequency (RF) spectrometer, to sound Pluto's atmosphere to derive its structure and composition. For either mode the electronics are unchanged from Radio Science Experiment (REX) on New Horizons (Tyler et al., 2008) but we note that this instrument requires an ultra-stable oscillator (USO) to provide a stable source (i.e., one with temperature control). The RF (radio-frequency) spectrometer is integrated into the RF communications system. It consists of electronics that monitor the RF transmissions from Earth while Pluto's atmosphere occults the signals. The resulting change in amplitude provides density and temperature information about a target's surface to the lowest few scale heights.

***Mass spectrometer:*** The mass spectrometer measures neutrals at densities $>10^4$ molecules $cm^{-3}$ and is based on the MAss SPectrometer for Planetary EXploration/Europa (MASPEX) instrument (Europa Clipper; e.g., Waite et al., 2019) but without the extra radiation shielding required for the Jovian environment and without the cold trap, which is unnecessary in the Pluto/Charon system. We also added an open aperture, which enables measurements of ions. Measurements up to 1000 μ are possible, but most neutral molecules would be below 200 μ. Mass resolution would be at least 2500 and is adjustable to 25,000.

***Altimeter:*** The altimeter is a version of the Mercury Laser Altimeter (MLA) modified to increase pulse rate from 8 Hz to 30 Hz (MESSENGER; Cavanaugh et al., 2007). It measures along-track topography at 1064 nm with high accuracy (range accuracy of 0.1 m) at 30 Hz, which corresponds to <30 m between measurements on the surface. The data are used to anchor elevations from stereo and to provide the shape accuracy required for inferring internal structure from gravity. The maximum range is at least 1200 km. The altimeter will be used to explore regions that are not sunlit (e.g. the southern pole of Pluto and Charon) during close-altitude passes (<1,200 km). In the nominal tour this will include five passes for Pluto's south pole, and two for Charon (c.f. Figure 15). These data can also be used to determine the albedo of the winter-dark poles along the tracks (akin to Qiao et al., 2019).

***Sounding radar:*** The sounding radar is a very-low-frequency (VLF), single-frequency (50 Hz) active radar based on SHARAD (Mars Reconnaissance Orbiter; Seu et al., 2007). This frequency is a balance between depth resolution of 3–5 m, depending on the layer composition, and the deepest penetration, which would be ~10 km. Pluto's ionosphere is too thin to interfere with the sounding radar measurements. The boom is two linear dipole antennas, and the radiated power is 100 W. The spatial resolution is ~4 km but can be shortened along-track to 1 km.

***Magnetometer:*** Two boom-mounted three-axis fluxgate magnetometers measure the magnetic field. The boom is 3.6 m (similar to the boom on the MESSENGER spacecraft) with one magnetometer mounted at the end and one approximately halfway down the boom. The instrument would operate continuously. The fluxgate magnetometers would be based on those on the Interstellar Mapping and Acceleration Probe (IMAP) (McComas et al., 2018). Two magnetometers provide a cleaner measurement by being able to compare the two, and offers redundancy, but the two are separate measurements (i.e. it's not the difference between the two that is required).

**Plasma spectrometer (CoDICE):** The plasma spectrometer measures the composition, energies, angular and spatial distributions, and densities of pickup ions, solar wind ions, and ionosphere ions. If a bow shock is present, the plasma spectrometer would quantify the densities. The energy range is between 3 eV and 50 keV. Both electrostatic analyzers and time-of-flight are used to determine composition. There are several existing instruments that can be flown. CoDICE-Lo on IMAP (McComas et al., 2018) is a recent example. A FIPS-type instrument (MESSENGER; Andrews et al., 2007) has a large field of view (FOV) that is well suited for non-spinning spacecraft. The instrument would operate when near the predicted bow shock and magnetotail and when within 3000 km of Pluto's surface. It would also gather data throughout cruise, the Pluto/Charon system tour and at during other KBO encounters to check for additional sources and flows.

## 4. Mission Design

The Persephone mission was designed to probabilistically encounter a KBO before entering and after departing Pluto orbit while accommodating the shortest technically feasible Earth-Pluto transfer time. The mission design is thus broken into three phases: the interplanetary transfer phase, the Pluto orbit phase, and the post-Pluto phase. Each phase is discussed in more detail below and Table 1 includes a timeline of major events throughout the mission.

**Table 1**

Persephone mission events table

| Major Event | Date | Time since launch (yrs) |
|---|---|---|
| Launch | February 2031 | 0.0 |
| Jupiter flyby | May 2032 | 1.2 |
| Pre-Pluto KBO flyby | February 2050 | 19.0 |
| Pluto arrival | October 2058 | 27.6 |
| Pluto departure | November 2061 | 30.7 |
| Post-Pluto KBO flyby | 2069-2090 | 38.7-59.7 |

*4.1. Interplanetary Transfer Phase*

The interplanetary transfer trajectory involves a high-energy launch on a Space Launch System (SLS) Block 2 with Centaur kick stage and Jupiter gravity assist en route to Pluto. Potential launch opportunities exist every year from 2029 to 2032, after which unfavorable geometry with Jupiter prevents subsequent launch opportunities until 2042. Performing a Jupiter gravity assist is required to deliver the required mass into Pluto orbit. Figure 9 shows the 21-day launch period that opens on 19 February 2031 with a maximum $C_3$ (characteristic energy, the excess specific energy after escaping from a body's gravity well) of 143 km$^2$ s$^{-2}$. The backup launch period occurs one year later in 2032. The Jupiter gravity assist is set to a minimum altitude of 17.8 Jupiter radii. In addition to the gravity assist, the trajectory accommodates a KBO flyby before arriving at the Pluto system.

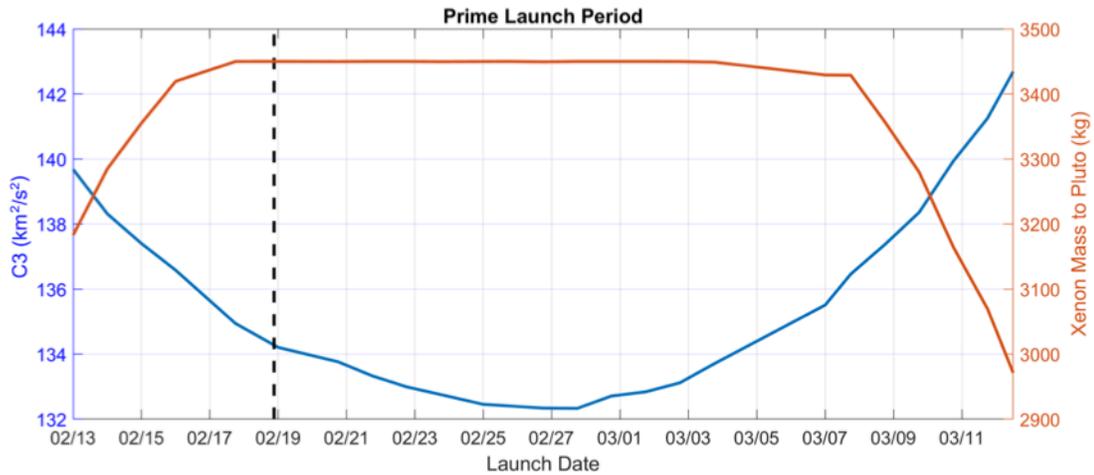

**Figure 9: Prime launch period (2031 February 13 to 2031 March 10) with corresponding C$_3$ (blue) and required xenon reaction mass (orange) as a function of time. The black dotted line shows the launch window opening date (19 February 2031).**

Statistical modeling was performed to identify the locations of KBOs along the nominal trajectory to Pluto. The model used was based on the Canada-France Ecliptic Plane Survey (CFEPS) studies of the inner, main, and outer classical belt and the resonant populations that have at least one member (Kavelaars et al., 2009; Petit et al., 2011; Gladman et al., 2012). The model was debiased to predict true orbital distributions to an $H_g$ magnitude of 8.5, and it includes the Centaur population. The assumption that all KBOs with a $H_g$ magnitude <8 will be known by the end of the 2020s is a reasonable one, since the Vera Rubin Observatory LSST is expected to catalog all such KBOs for a distance of <45 AU by 2027. The Nancy Grace Roman Space Telescope is also expected to be operational by the late 2020s, which will assist with this effort.

The model results are shown in Figure 10 and Figure 11, and the assumed relationship between brightness and object size is given in Figure 12. These results indicate that a 0.2 AU deviation from the nominal trajectory would enable a flyby of ~30-km-sized KBO (i.e., approximately the same size as Arrokoth), and a >0.5-AU deviation would be required to visit a larger target. Analysis of the trade-off between the time-of-flight penalty of a trajectory deviation and the magnitude of the deviation itself (Figure

13) led us to assume a 1-AU deviation, which would add one Earth-year to Persephone's cruise time but would enable a ~50- to 100-km-sized KBO to be visited.

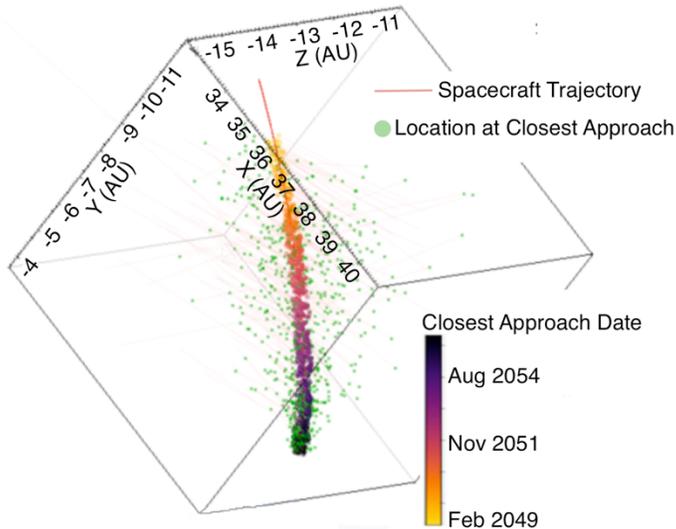

**Figure 10: Distance and closest approach date of Persephone to the modeled KBO population, for the pre-Pluto encounter. Close approach dates range from 2049-2054, 18-23 years after the optimal launch date of 2031.**

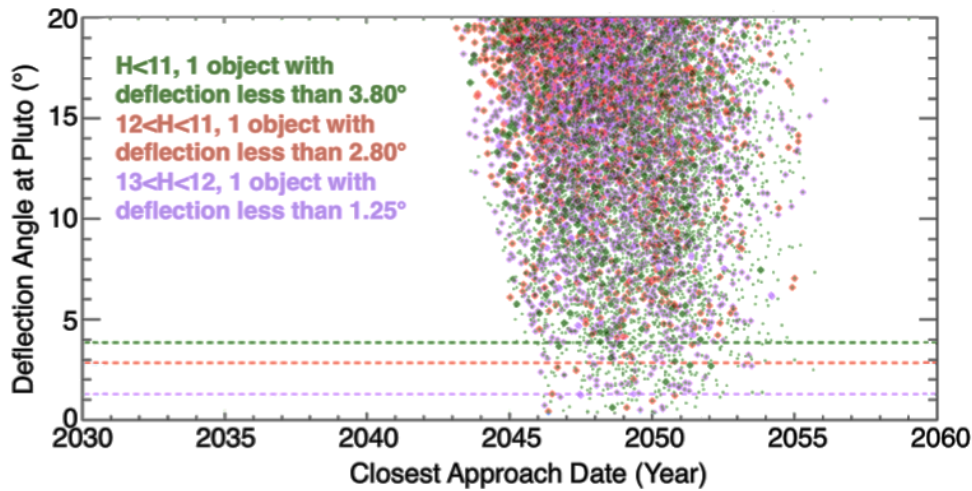

**Figure 11: Deflection angle at Pluto (i.e., how modified the nominal trajectory would have to be) to encounter KBOs of different brightness. Intrinsically fainter (smaller) targets are more numerous and more easily reached at smaller deflection angles.**

*4.2. Pluto Orbit Phase*

Due to the premium placed on the mass required to deliver Persephone into Pluto orbit, the science orbit consists of multiple periodic orbits designed in the Pluto/Charon restricted three-body dynamics model. These orbits leverage the simultaneous gravitational pull from both Pluto and Charon and effectively eliminate the need for high-thrust chemical propulsion maneuvers during the Pluto orbit phase (a small amount is allocated for station-keeping). The Pluto/Charon system, with mass proportions an order of magnitude greater than the Earth/Moon system, gives an unprecedented opportunity to exploit restricted three-body dynamics to find very chaotic orbits that span very large radial distances both in and out of the satellite plane, enabling close encounters with Pluto and all its satellites. These orbits, when viewed in a Pluto-centered inertial frame, do not follow standard Keplerian motion because of gravitational perturbations from Charon, but, when viewed in a frame that has both Pluto and Charon fixed, can reveal insightful characteristics. The periodic nature of these orbits in the Pluto/Charon rotating frame, coupled with the fact that Pluto and Charon are tidally locked, have repeating ground tracks naturally built into the trajectory. The periodic orbits identified fell into two primary categories: a high out-of-plane component to enable high-latitude global mapping and low altitude to enable in situ sampling. Four distinct periodic orbits—two from each category—consisting of the complete Pluto orbit tour were selected (Figures 14 and 15). The orbits also have varying maximum radial extents from Pluto, enabling encounter opportunities with the smaller moons. Persephone would have the capability to transfer between periodic orbits. The total science orbit mission duration, including time spent performing multiple revolutions on certain periodic orbits and the transfers between orbits, is 3.1 years.

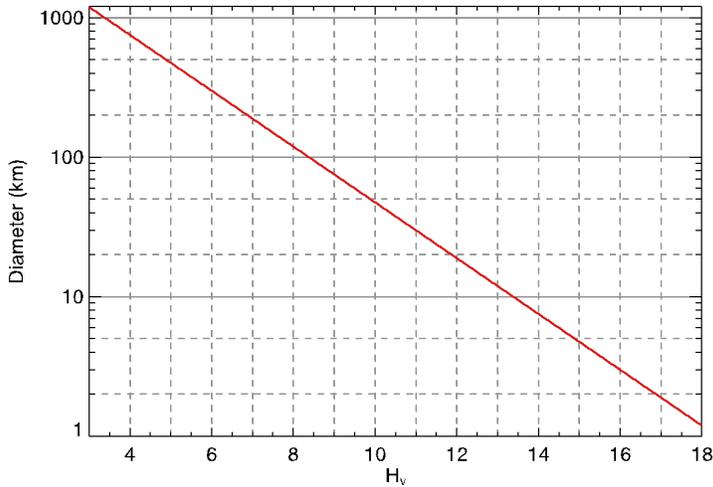

**Figure 12: Relationship between the absolute magnitude, $H_v$, and the diameter of a body, assuming a visible geometric albedo, $p_V$, of 0.08, an average value for KBOs. This equation governing this relationship was determined by Harris (1998).**

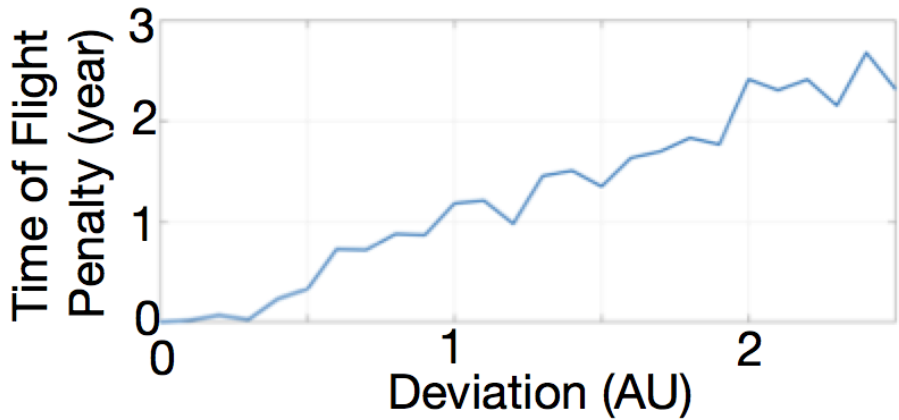

**Figure 13: Time-of-flight penalty for nominal trajectory deviations between 0 and 2.5 AU for a pre-Pluto KBO encounter. As shown in Figure 11, larger deviations enable encounters with larger KBOs.**

While the Pluto-Charon tour is relatively insensitive to the arrival time at the Pluto system, this is not true for the minor satellites. Once the arrival time is defined, the mission would be able to define the minor satellite tour by seeing when the closest approaches are to each satellite. Generally, the closer a minor satellite is to Pluto, the more and closer potential encounters there are with it. Thus, the best coverage would likely be obtained for Styx, and the worst for Hydra. Depending on the timing, it may be possible to have a Kerberos or Hydra close flyby during the transition period between orbital configurations. The encounter velocity of the minor satellites would be <300 m s⁻¹.

Persephone's expected (and New Horizons' achieved) resolution for the minor satellites would be (Weaver et al., 2016):

- Styx: 80 m/pixel (3157 m/pixel)
- Nix: 80 m/pixel (306 m/pixel)
- Kerberos: 110 m/pixel (1982 m/pixel)
- Hydra: 175 m/pixel (1155 m/pixel)

In addition, Persephone would be able to image the minor satellites at additional geometries that were not achievable during the New Horizons flyby. Persephone would almost certainly be able to expand the global imaging coverage for all of the minor satellites.

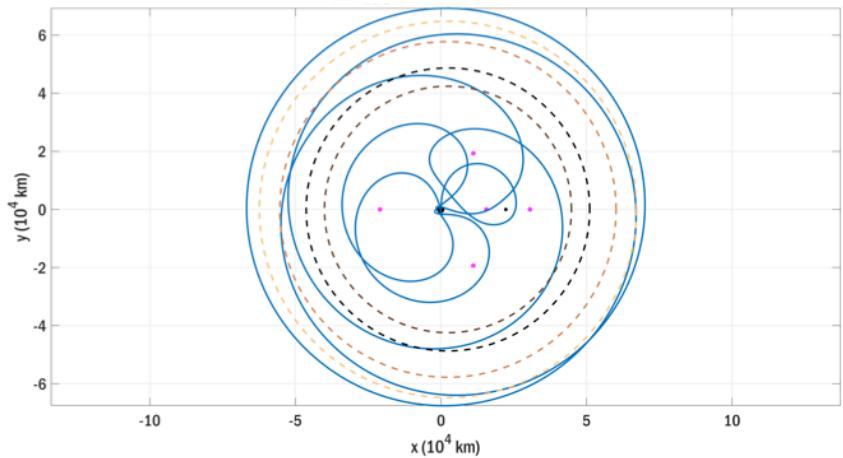

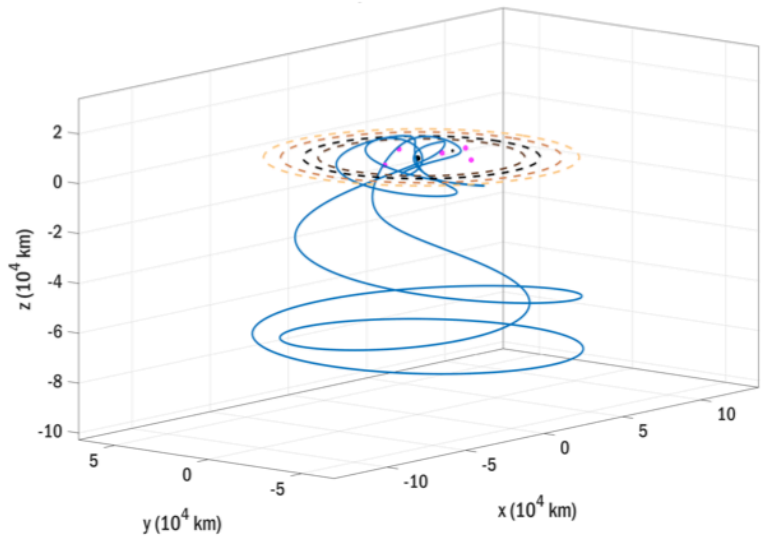

Persephone: Pluto Orbiter

**(a) Orbit 1**

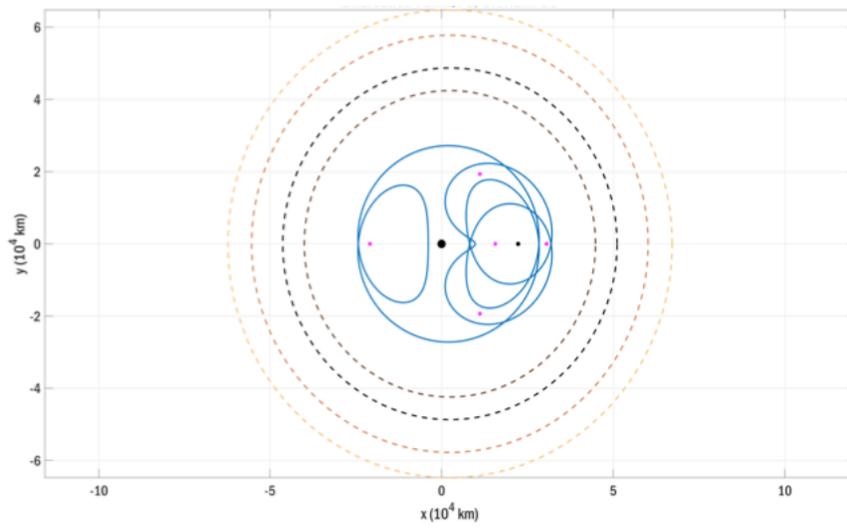

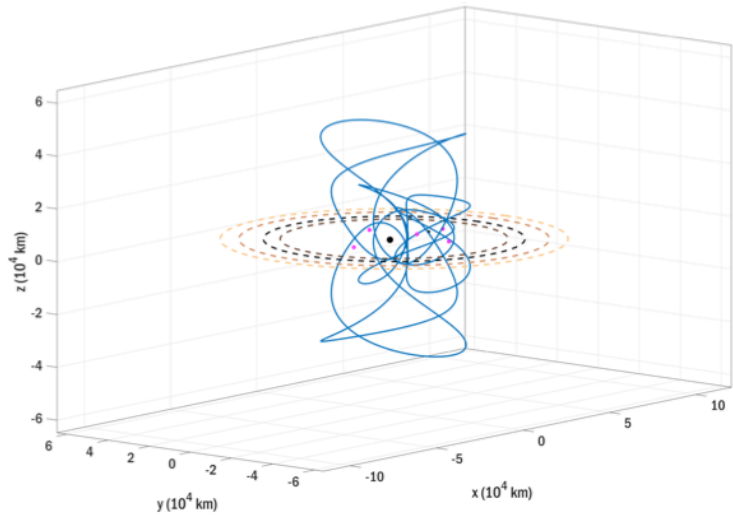

**(b) Orbit 2**

Persephone: Pluto Orbiter

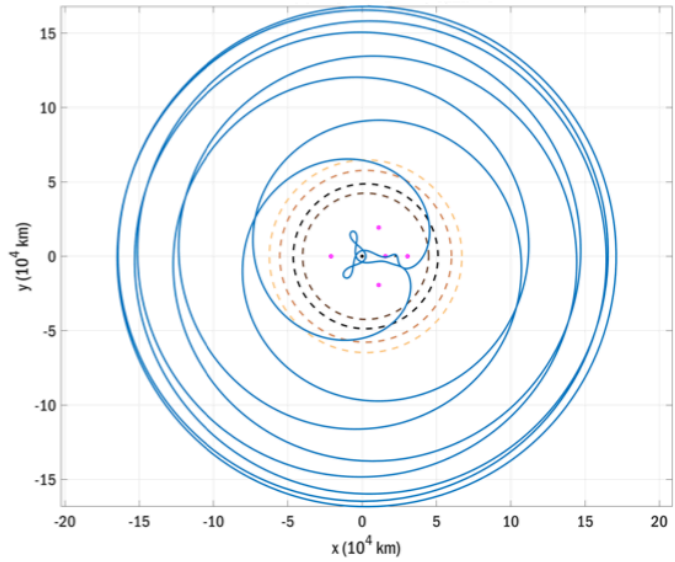

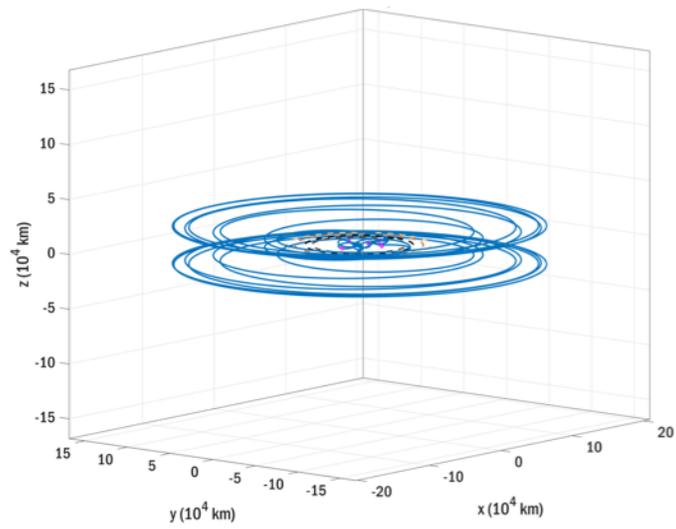

**(c) Orbit 3**

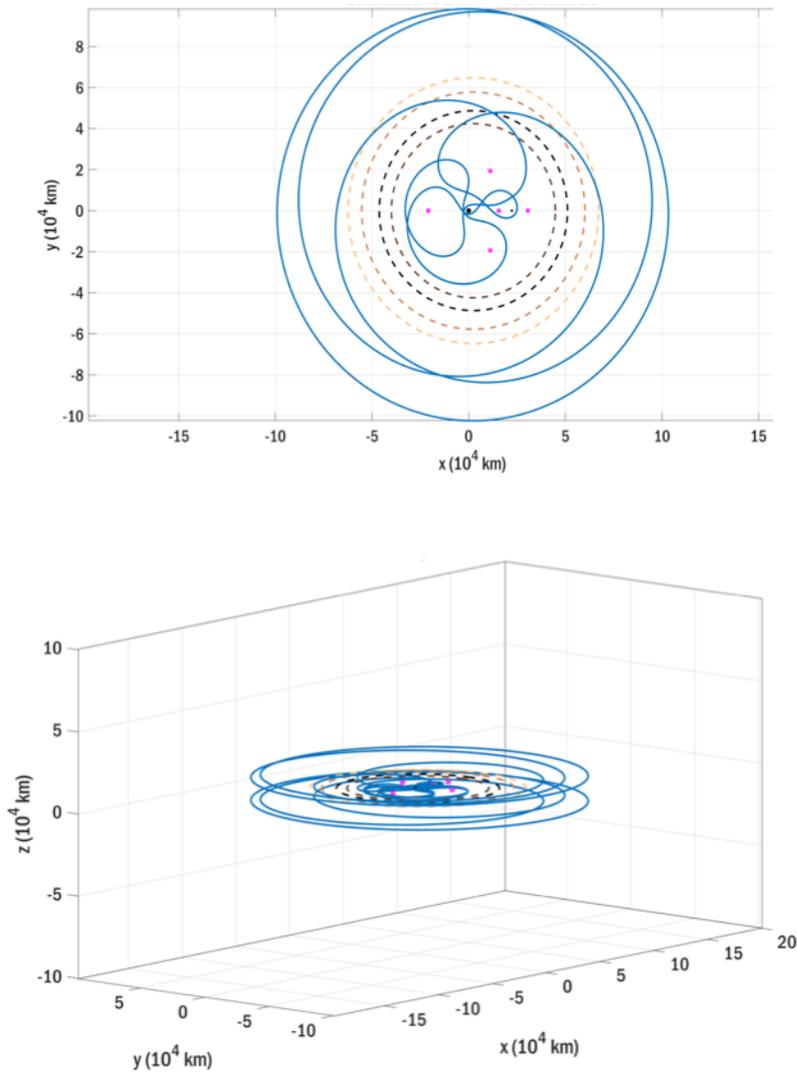

**(d) Orbit 4**

**Figure 14: The geometry of the science orbits, plotted in the Pluto/Charon rotating frame from two different vantage points. The dotted lines indicate the orbits of Pluto's minor satellites, the black dots show the location of Pluto and Charon, and the pink dots indicate the Lagrange points. The four chosen orbits are shown by (a) to (d). Axes are distance from Pluto (in $10^4$ km).**

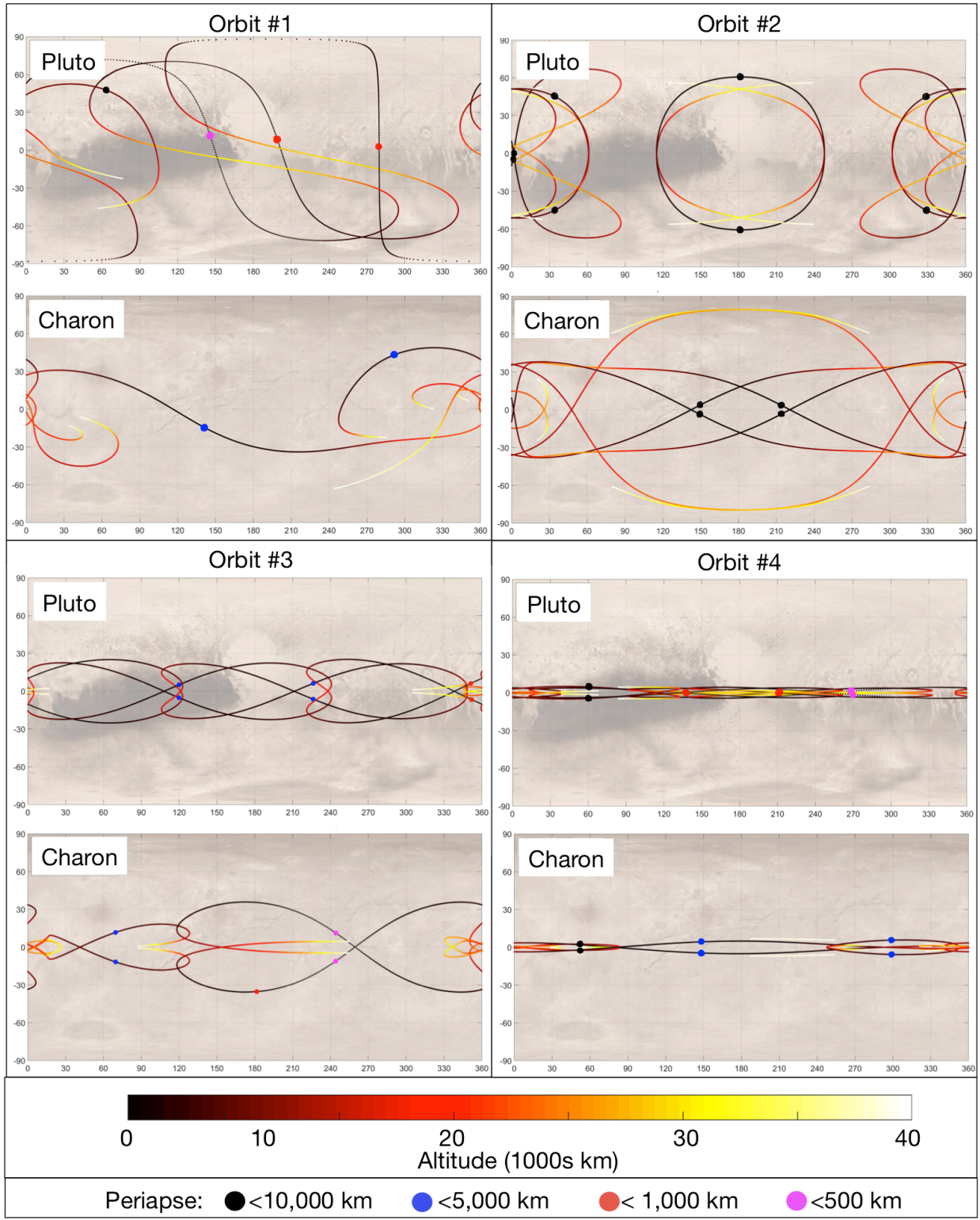

**Figure 15: Global coverage of Pluto and Charon provided by each of the four chosen orbits. The color indicates the altitude (see color bar), and the ground-**

**track locations at periapses are depicted as colored dots, with the color indicative of altitude (see subfigure keys for details). Orbits 1 and 2 provide global coverage of both targets, while Orbits 3 and 4 provide low-altitude and low-latitude coverage that is optimized for in situ instruments.**

*4.3. Post-Pluto Phase*

As a potential extended mission option, a Charon gravity assist can be used to depart the Pluto system. Xenon propellant available at the end of the Pluto orbit phase limits Pluto departure velocity and thus transfer time to more distant KBOs (Figure 16). Robust amounts of xenon margin in the mission propellant budget could accommodate access to many targets. For example, using 250 kg of xenon from margin enables reaching a target 3 AU from Pluto after ~8 years, which based on analysis of KBO statistical position models could be a flyby of a 100- to 150-km target.

Similar modeling to the pre-Pluto KBO encounter was done for the post-Pluto encounter as well. There is a clear trade-off between the post-Pluto distance traveled, time, and xenon used. A 1 (3) AU deviation would enable an object ~95 km (200 km) diameter to be visited, this corresponds to objects with $H_V < 9$ (<7). Such an extended mission would require additional xenon, which would have to be carried throughout the entire nominal mission, affecting the initial cruise duration. We selected 250 kg of xenon as a compromise to enable a 100-km statistical object to be reached in ~6 years. Additionally, a list of possible known KBO targets, and the time needed to reach them, is given in Table 2.

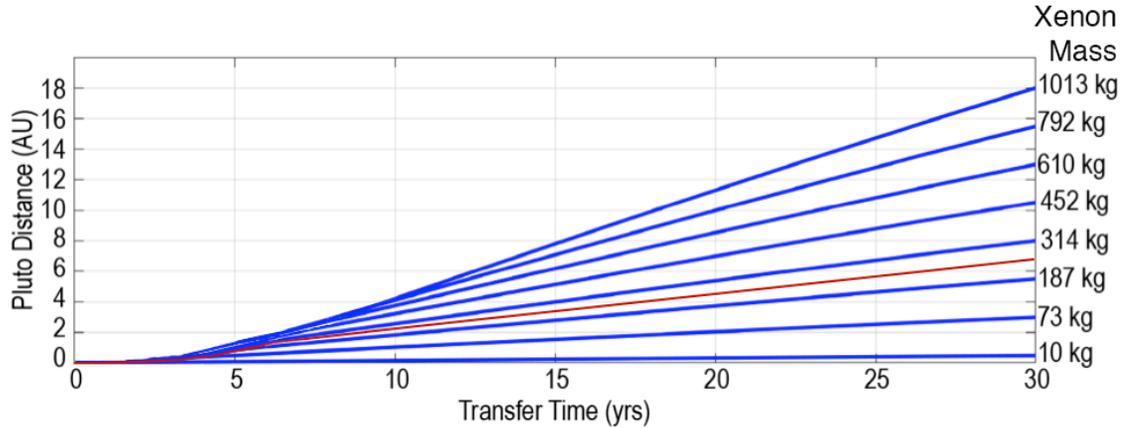

**Figure 16: Post-Pluto transfer time and the required xenon reaction mass (as indicate on the right hand side of the figure) to achieve the science goal of encountering a 100- to 150-km-class KBO. The 250 kg reserved for the extended mission is shown as the red line that lies approximately halfway between the blue curves representing 187 kg and 314 kg.**

**Table 2**

List of known post-Pluto KBO targets that could be reached

| Name | $H_V$ | Est. diam. (km)[a] | Approx. departure | Transit duration (yrs) | Dist. Range (AU) |
|---|---|---|---|---|---|
| (470308) 2007 JH43 | 4.5 | 600 | 2061–2066 | 20 | 4.39-4.46 |
| (182294) 2001 KU76 | 6.6 | 230 | 2060–2066 | 17–20 | 3.02-4.37 |
| 2004 HN79 | 7.1 | 180 | 2060–2066 | 18–20 | 2.63-4.27 |
| 2004 HZ78 | 7.3 | 165 | 2060–2061 | 16–19 | 3.37-4.29 |
| 2015 KD176 | 7.4 | 160 | 2060–2070 | 20 | 3.93-4.05 |
| (523704) 2014 HB200 | 7.5 | 150 | 2060–2061 | 17–20 | 3.61-4.27 |
| 2013 JR65 | 7.9 | 125 | 2060–2067 | 13–17 | 2.26-3.73 |
| 2000 FT53 | 8.3 | 100 | 2060–2066 | 17–20 | 2.09-4.35 |
| 2004 HY78 | 8.3 | 100 | 2060–2064 | 15–20 | 2.55-4.48 |
| 2015 KO173 | 8.8 | 80 | 2060–2070 | 12–19 | 0.41-4.26 |
| 2015 KQ175 | 8.9 | 75 | 2060–2070 | 20 | 3.60-3.83 |
| 2015 GH54 | 9.7 | 55 | 2060–2070 | 18–20 | 1.47-4.04 |

[a]Assuming a visible geometric albedo, $p_V$, of 0.08 and the relation of Harris (1998). See Figure 12. **Altenhoff et al. (2004) showed that geometric albedos of KBOs varied from 0.06 to 0.16 but the average value was 0.08, which is why this number is adopted here. This albedo is consistent with that found by Lacerda et al. (2014).**

## 5. Summary

Persephone is an ambitious mission concept that would enter orbit in the Pluto system and explore the surface compositions, atmospheric structure, geologic history, and radiation environment of Pluto, Charon, and its minor satellites in unprecedented detail after an unprecedented transfer time of several decades. Planned fuel reserves would enable encounters with other KBOs prior to and following the Pluto orbit phase, shedding light on the larger KBO population. A suite of 11 instruments capable of imaging, UV and IR spectroscopy, mass spectroscopy, plasma spectroscopy, radar sounding, radio science, magnetic field measurements, and altimetry would be brought to bear on any and every object encountered during the mission. Persephone was designed with redundancy and longevity in mind, given a transit time to Pluto of 27.6 years following launch on an SLS Block 2 and a Jupiter gravity assist, 3.1 years in the Pluto system executing 4 different repeating orbits to optimize coverage, and up to 20 years for an extended mission to an intermediate-sized KBO. In all, this mission could operate for over 50 years, challenging engineering, mission operations, and data analysis in ways that have never been done before. The wealth of knowledge delivered during the Persephone mission would far outlast the mission lifetime, providing several generations of scientists an unparalleled look at the Pluto system and the Kuiper Belt.

## Acknowledgements

We thank NASA for its financial support through the Planetary Mission Concept Studies (PMCS) grant #18-PMCS18-0027 and contract task NNN06AA01C/80MSFC19F0097. We also thank SwRI and APL for their financial support through internal research funding (SwRI #R6007).